\documentclass[11pt]{article}
\usepackage{amsmath,amsthm,amscd,array,amssymb}

\begin{document}

\begin{center}
{\Large{\bf Cohomological extension of $Spin(7)$--invariant\\
\medskip 
super Yang--Mills theory in eight dimensions}}
\\
\bigskip\medskip
{\large{\sc D. M\"ulsch}}$^{a}$
\footnote{Email: muelsch@informatik.uni-leipzig.de}
and
{\large{\sc B. Geyer}}$^b$
\footnote{Email: geyer@itp.uni-leipzig.de}
\\
\medskip
{\it $\!\!\!\!\!^a$ Wissenschaftszentrum Leipzig e.V., D--04103 Leipzig, Germany
\\
\smallskip
$^b$ Universit\"at Leipzig, Naturwissenschaftlich-Theoretisches Zentrum
\\
$~$ and Institut f\"ur Theoretische Physik, D--04109 Leipzig, Germany}
\end{center}
\smallskip

\begin{abstract}
\noindent {\small{It is shown that the $Spin(7)$--invariant super Yang--Mills 
theory in eight dimensions, which relies on the existence of the Cayley 
invariant, permits the construction of a cohomological extension,
which relies on the existence of the eight--dimensional analogue of the
Pontryagin invariant arising from a quartic chiral primary operator.}}
\end{abstract}


\section{Introduction}

Topological quantum field theory (TQFT) has attracted a lot of
interest over the last years, both for its own sake and due to
their connection with string theory (for a review, see, e.g., \cite{1}). 
Particularly interesting are TQFT's in $D = 2$ \cite{2}, because of their 
connection with $N = 2$ superconformal theories and with Calabi--Yau moduli 
spaces \cite{3}. Considerable impact on physics and mathematics has 
had Witten's construction of topological Yang--Mills theory in $D = 4$
and the discovery of its relation with the Donaldson map, which
relates the de Rham cohomology groups on four--manifolds with those
on the moduli space of quaternionic instantons, as well as its relation 
to the topologically twisted super Yang--Mills theory \cite{4}.

Recently, the construction of cohomological gauge theories on manifolds 
of special holonomy in $D > 4$ have received considerable attention, 
too \cite{5,6,7}. These theories, which arise without the necessity for 
a topological twist, acquire much of the characteristics of a TQFT. 
However, such theories are not fully topological, since they are only 
invariant under such metric variations which do not change the reduced 
holonomy structure. For $D = 8$ examples of cohomological gauge theories
have been constructed for the cases when the holonomy group in
$SO(8)$ is either $Spin(7)$ \cite{5,6} (Joyce manifolds) or
$Spin(6) \sim SU(4)$ \cite{5} (Calabi--Yau four--folds) or 
$Spin(5) \sim Sp(4)$ \cite{8} (hyper--K\"ahler eight--folds).

At present, the physical content of these theories has not been
entirely revealed, especially, since they are not renormalizable.
But, recent developments of string theory have renewed the
interest in super Yang--Mills theories (SYM) in $D > 4$, particulary
because of their crucial role in the study of D--branes and in the
matrix approach to M--theory. It is widely believed that the
low--energy effective world volume theory of D--branes obtains
through dimensional reduction of $N = 1$, $D = 10$ SYM \cite{9}. 
Hence, the above mentioned cohomological theories in $D = 8$ provide 
effective field theories on the world volume of Euclidean Dirichlet 
7--branes wrapping around manifolds of special holonomy. 
In order to improve their renormalizibility, counterterms are needed 
at high energies arising from string theory compactifications down to 
eight dimensions. Such terms were computed in \cite{10}. In \cite{11} 
it was argued that the complete string--corrected counterterms for the
$Spin(7)$--invariant theory, whose construction relies on the existence 
of the Cayley invariant, should be still cohomological. 

In this paper we show --- without going beyond the scope of a cohomological 
theory --- that the $Spin(7)$--invariant theory, in fact, has a cohomological 
extension. It relies on the existence of the eight--dimensional analogue 
of the Pontryagin invariant which arises from the primary operator 
$W_0 = \frac{1}{4}\, {\rm tr}\, \phi^4$. This extension, which in flat
space is uniquely determined by the shift and vector supersymmetries, 
can be related to the one--loop string--corrected counterterms.

The paper is organized as follows: In Sect. 2, we briefly describe
the formulation of the $Spin(7)$--invariant $N_T = 1$, $D = 8$ SYM. 
By analyzing the structure of the observables it is argued that this theory 
permits the construction of a cohomological extension which relies 
on the existence of the eight--dimensional analogue of the Pontryagin 
invariant. 
In Sect. 3, by performing a dimensional reduction to $D = 4$, it is shown 
that the matter--independent part of the resulting half--twisted theory 
\cite{12} can be obtained from the operator 
$\hat{W}_0 = \frac{1}{2}\, {\rm tr}\, \phi^2$ by means of a relationship 
associated with the vector supersymmetry. 
In Sect. 4, by generalizing this relationship to $D = 8$, we give the
cohomological extension of the $Spin(7)$--invariant theory in the
Landau type gauge. 
Appendix A contains the Euclidean spinor conventions which are used. 
Appendix B lists the off--shell transformation rules for the 
matter--independent part of the half--twisted theory. 
Appendix C gives, in some detail, the derivation of the cohomological 
extension in the Feynman type gauge.

\section{ $Spin(7)$--invariant, $D = 8$ Euclidean super Yang--Mills theory}

An eight--dimensional analogue of Donaldson--Witten theory on $Spin(7)$ 
holonomy Joyce manifold, which localizes onto the moduli space of octonionic 
instantons, has been constructed in \cite{5,6}. This theory, just as 
Donaldson--Witten theory on $SU(2)$ holonomy hyper--K\"ahler two--fold, 
is topological without twisting. In fact, it is invariant under metric 
variations preserving the $Spin(7)$ structure. In flat space the theory 
arises from the $N = 2$, $D = 8$ Euclidean SYM by reducing the $SO(8)$ 
rotation invariance to $Spin(7)$. This reduction has been explicitly 
carried out in \cite{13} (preserving hermiticity and imposing some extra 
constraints) and in \cite{14} (relaxing the reality conditions on fermions 
without introducing extra constraints). Thereby, extensive use is made 
of the octonionic algebra together with their compatibility with 
$Spin(7)$ invariance, generalized self--duality and chirality.

The action of the $Spin(7)$--invariant $N_T = 1$, $D = 8$ Euclidean SYM 
is built up from the 10 bosonic scalar and vector fields
$\phi$, $\bar{\phi}$ and $A_a$ ($a = 1, \ldots, 8$), respectively, and
from the 16 fermionic scalar, vector and self--dual tensor fields $\eta$,
$\psi_a$ and $\chi_{ab} = \frac{1}{6} \Phi_{abcd} \chi^{cd}$,
respectively. $\Phi_{abcd}$ is the $Spin(7)$--invariant, completely 
antisymmetric self--dual Cayley tensor, $\Phi_{abcd} =
\frac{1}{24} \epsilon_{abcdefgh} \Phi^{efgh}$, with $\epsilon_{abcdefgh}$ 
being the Levi--Civita tensor in eight dimensions. All the fields take 
their values in the Lie algebra $Lie(G)$ of some compact gauge group $G$. 
Adopting the notation of \cite{14} the $Spin(7)$--invariant action reads
\begin{align}
\label{2.1} 
S = \int_E d^8x\, {\rm tr} \Bigr\{&
\hbox{$\frac{1}{8}$} \Theta^{abcd} F_{ab} F_{cd} - 2 D^a
\bar{\phi}\, D_a \phi - 2 \chi^{ab} D_a \psi_b + 2 \eta D^a \psi_a
\nonumber
\\
& + 2 \bar{\phi} \{ \psi^a, \psi_a \} +
\hbox{$\frac{1}{4}$} \phi \{ \chi^{ab}, \chi_{ab} \} +
2 \phi \{ \eta, \eta \} - 2 [ \phi, \bar{\phi} ]^2 \Bigr\},
\end{align}
where $F_{ab} = \partial_{[a} A_{b]} + [ A_a, A_b ]$ and
$D_a = \partial_a + [ A_a, ~\cdot~ ]$. Here, the (unnormalized) 
projector \cite{6}
\begin{equation*}
\Theta_{abcd} = \delta_{ac} \delta_{bd} - \delta_{ad} \delta_{bc} +
\Phi_{abcd},
\qquad
\hbox{$\frac{1}{8}$} \Theta_{abef} \Theta_{cd}^{~~~ef} = \Theta_{abcd},
\end{equation*}
projects any antisymmetric second rank tensor onto its self--dual
part $\mathbf{7}$ according to the decomposition
$\mathbf{28} = \mathbf{7} \oplus \mathbf{21}$ of the adjoint representation 
of $SO(8) \sim SO(8)/Spin(7) \otimes Spin(7)$. It satisfies the 
relations \cite{14}
\begin{align}
\label{2.2}
&\hbox{$\frac{1}{2}$} ( \Theta_{abeg} \Theta_{cdf}^{~~~~\!g} -
\Theta_{abfg} \Theta_{cde}^{~~~~\!g} ) =
- \Theta_{efac} \delta_{bd} + \Theta_{efad} \delta_{bc} +
\Theta_{efbc} \delta_{ad} - \Theta_{efbd} \delta_{ac}
\nonumber
\\
&\phantom{\hbox{$\frac{1}{2}$} ( \Theta_{abeg} \Theta_{cdfg} -
\Theta_{abfg} \Theta_{cdeg} ) =,}
+ \Theta_{abce} \delta_{df} - \Theta_{abde} \delta_{cf} -
\Theta_{abcf} \delta_{de} + \Theta_{abdf} \delta_{ce}
\nonumber
\\
&\phantom{\hbox{$\frac{1}{2}$} ( \Theta_{abeg} \Theta_{cdfg} -
\Theta_{abfg} \Theta_{cdeg} ) =,}
- \Theta_{cdae} \delta_{bf} + \Theta_{cdbe} \delta_{af} +
\Theta_{cdaf} \delta_{be} - \Theta_{cdbf} \delta_{ae},
\nonumber
\\
&\hbox{$\frac{1}{2}$} ( \Theta_{abeg} \Theta_{cdf}^{~~~~\!g} +
\Theta_{abfg} \Theta_{cde}^{~~~~\!g} ) =
\Theta_{abcd} \delta_{ef},
\end{align}
enclosing all the properties of the structure constants entering the 
octonionic algebra \cite{15}.

On--shell, upon using the equation of motion for $\chi^{ab}$, the action
(\ref{2.1}) can be recast into the $Q$--exact form, $S = Q \Psi$, where the
gauge fermion
\begin{equation}
\label{2.3}
\Psi = \int_E d^8x\, {\rm tr} \Bigr\{
\chi^{ab} ( F_{ab} - \hbox{$\frac{1}{16}$} \Theta_{abcd} F^{cd} ) -
2 \psi^a D_a \bar{\phi} - 2 [ \eta, \bar{\phi} ] \phi \Bigr\}
\end{equation}
is uniquely fixed by requiring its invariance under vector supersymmetry, 
$Q_a \Psi = 0$.

The full set of supersymmetry transformations, generated by the supercharges 
$Q$, $Q_a$ and $Q_{ab} = \frac{1}{6} \Phi_{abcd} Q^{cd}$, which leave the 
action (\ref{2.1}) invariant, are given by \cite{14},
\begin{alignat}{2}
\label{2.4}
&Q A_a = \psi_a,
&\qquad
&Q \psi_a = D_a \phi,
\nonumber
\\
&Q \phi = 0,
&\qquad
&Q \bar{\phi} = \eta,
\nonumber
\\
&Q \eta = [ \bar{\phi}, \phi ],
&\qquad
&Q \chi_{ab} = \hbox{$\frac{1}{4}$} \Theta_{abcd} F^{cd},
\\
\nonumber
\\
\label{2.5}
&Q_a A_b = \delta_{ab} \eta + \chi_{ab},
&\qquad
&Q_a \psi_b = F_{ab} - \hbox{$\frac{1}{4}$} \Theta_{abcd} F^{cd} +
\delta_{ab} [ \phi, \bar{\phi} ],
\nonumber
\\
&Q_a \phi = \psi_a,
&\qquad
&Q_a \bar{\phi} = 0,
\nonumber
\\
&Q_a \eta = D_a \bar{\phi},
&\qquad
&Q_a \chi_{cd} = \Theta_{abcd} D^b \bar{\phi}
\\
\intertext{and}
\label{2.6}
&Q_{ab} A_c = - \Theta_{abcd} \psi^d,
&\qquad
&Q_{ab} \psi_c = \Theta_{abcd} D^d \phi,
\nonumber
\\
&Q_{ab} \phi = 0,
&\qquad
&Q_{ab} \bar{\phi} = \chi_{ab},
\nonumber
\\
&Q_{ab} \eta = - \hbox{$\frac{1}{4}$} \Theta_{abcd} F^{cd},
&\qquad
&Q_{ab} \chi_{cd} = \hbox{$\frac{1}{4}$} \Theta_{abeg}
\Theta_{cdf}^{~~~~\!g} F^{ef} + \Theta_{abcd} [ \bar{\phi}, \phi ].
\end{alignat}
The {\it on--shell} algebra of the supercharges $Q$, $Q_a$ and $Q_{ab}$ reads
\begin{align}
\label{2.7}
&\{ Q, Q \} \doteq - 2 \delta_G(\phi),
\qquad
\{ Q, Q_a \} \doteq \partial_a + \delta_G(A_a),
\qquad
\{ Q_a, Q_b \} \doteq - 2 \delta_{ab} \delta_G(\bar{\phi}),
\nonumber
\\
&\{ Q_{ab}, Q \} \doteq 0,
\qquad
\{ Q_{ab}, Q_c \} \doteq \Theta_{abcd} ( \partial^d + \delta_G(A^d) ),
\qquad
\{ Q_{ab}, Q_{cd} \} \doteq - 2 \Theta_{abcd} \delta_G(\phi),
\end{align}
where $\delta_G(\varphi)$ denotes a gauge transformation with
field--dependent parameter $\varphi = ( A_a, \phi, \bar{\phi} )$,
being defined by $\delta_G(\varphi) A_a = - D_a \varphi$ and
$\delta_G(\varphi) X = [ \varphi, X ]$ for all the other fields.
(The symbol $\doteq$ means that the corresponding relation is
fulfilled only {\it on--shell}.)

The crucial hint, supporting our claim that (\ref{2.1}) allows for a
cohomological extension, comes from the structure of the topological 
observables. Namely, in terms of differential forms, the method of descent 
equation implies, starting from the primary operator 
$\tilde{W}_0 = \frac{1}{2}\, {\rm tr}\, \phi^2$, the following ladder 
of $k$--forms $\tilde{W}_k$ ($4 \leq k \leq 8$) with ghost number 
$8 - k$ \cite{5,6},
\begin{align}
\label{2.8}
&\tilde{W}_4 = \Phi \wedge {\rm tr}\, ( \hbox{$\frac{1}{2}$} \phi^2 ),
\nonumber
\\
&\tilde{W}_5 = \Phi \wedge {\rm tr}\, ( \psi \phi ),
\nonumber
\\
&\tilde{W}_6 = \Phi \wedge {\rm tr}\, ( F \phi +
\hbox{$\frac{1}{2}$} \psi \wedge \psi ),
\nonumber
\\
&\tilde{W}_7 = \Phi \wedge {\rm tr}\, ( F \wedge \psi + \psi \wedge \psi ),
\nonumber
\\
&\tilde{W}_8 = \Phi \wedge {\rm tr}\, ( \hbox{$\frac{1}{2}$} F \wedge F ),
\end{align}
with $\psi = \psi_a e^a$, $F = \frac{1}{2} F_{ab}\, e^a \wedge e^b$ and
$\Phi = \frac{1}{24} \Phi_{abcd}\, e^a \wedge e^b \wedge e^c
\wedge e^d$ ($e^a = e_\mu^{\!~~a} dx^\mu$, where $e_\mu^{\!~~a}$ is the 
8--bein on the $Spin(7)$--holonomy Joyce manifold $\cal{M}$ endowed with 
metric $g_{\mu\nu}$). The trace is taken in $Lie(G)$. These $k$--forms 
obey the following decent equations,
\begin{equation*}
0 = Q 
\tilde{W}_4,
\qquad
d \tilde{W}_k = Q \tilde{W}_{k + 1},
\qquad
4 \leq k \leq 7,
\qquad
d \tilde{W}_8 = 0,
\end{equation*}
which are typical for any cohomological gauge theory.

Hence, if $\gamma$ is a $k$--dimensional homology cycle on $\cal M$ then the
integrated descendants $\tilde{I}_k(\gamma) = \int_{\gamma} \tilde{W}_k$ 
($4 \leq k \leq 7$) and $\tilde{I}_8 = \int_{\cal M} \tilde{W}_8$ are 
$Q$--invariant, $Q \tilde{I}_k(\gamma) = \int_\gamma Q \tilde{W}_k = 
\int_\gamma d \tilde{W}_{k - 1} = 0$.
They depend, up to a $Q$--exact term, only upon the homology class of 
$\gamma$, i.e., when adding to $\gamma$ a boundary term $\partial \alpha$, 
then $\tilde{I}_k(\gamma)$ remains unaltered, modulo a $Q$--exact term,
$\tilde{I}_k(\gamma + \partial \alpha) = 
\int_{\gamma + \partial \alpha} \tilde{W}_k =
\tilde{I}_k(\gamma) + \int_\alpha d \tilde{W}_k = 
\tilde{I}_k(\gamma) + \int_\alpha Q \tilde{W}_{k + 1} = \tilde{I}_k(\gamma)$.
Here, the integrated descendant $\tilde{I}_8$ is the Cayley invariant, 
being invariant under a certain class of metric variations which do not 
change the reduced $Spin(7)$ structure.

By means of the same method, starting from the primary operator 
$W_0 = \frac{1}{4}\, {\rm tr}\, \phi^4$, one can derive the following 
ladder of $k$--forms $W_k$ ($0 \leq k \leq 8$) with ghost number $8 - k$,
\footnote{In order to condense the notation, we introduced a {\it symmetrized}
trace, ${\rm tr}_{\rm (s)}$, which is defined in Appendix C.}
\begin{align}
\label{2.9}
&W_0 = {\rm tr}_{\rm (s)}\, ( \hbox{$\frac{1}{4}$} \phi^4 ),
\nonumber
\\
&W_1 = {\rm tr}_{\rm (s)}\, ( \psi \phi^3 ),
\nonumber
\\
&W_2 = {\rm tr}_{\rm (s)}\, ( F \phi^3 +
\hbox{$\frac{1}{2}$} \psi \wedge \psi \phi^2 ),
\nonumber
\\
&W_3 = {\rm tr}_{\rm (s)}\, ( F \wedge \psi \phi^2 +
\psi \wedge \psi \wedge \psi ),
\nonumber
\\
&W_4 = {\rm tr}_{\rm (s)}\, ( \hbox{$\frac{1}{2}$} F \wedge F \phi^2 +
F \wedge \psi \wedge \psi \phi +
\hbox{$\frac{1}{4}$} \psi \wedge \psi \wedge \psi \wedge \psi ),
\nonumber
\\
&W_5 = {\rm tr}_{\rm (s)}\, ( \hbox{$\frac{1}{2}$} F \wedge F \wedge \psi \phi +
F \wedge \psi \wedge \psi \wedge \psi ),
\nonumber
\\
&W_6 = {\rm tr}_{\rm (s)}\, ( F \wedge F \wedge F \phi +
\hbox{$\frac{1}{2}$} F \wedge F \wedge \psi \wedge \psi ),
\nonumber
\\
&W_7 = {\rm tr}_{\rm (s)}\, ( F \wedge F \wedge F \wedge \psi ),
\nonumber
\\
&W_8 = {\rm tr}_{\rm (s)}\, ( \hbox{$\frac{1}{4}$} F \wedge F \wedge F \wedge F ),
\end{align}
which obey similar descendant equations as before. Again, we observe 
that the integrated descendant $I_8 = \int_{\cal M} W_8$ yields a topological 
invariant which is now unchanged under arbitrary metric variations and 
which may be regarded as the eight--dimensional analogue of the Pontryagin
invariant. This suggests that the $Spin(7)$--invariant theory permits, 
in fact, a cohomological extension, $S_{\rm ext} = Q \Psi_{\rm ext}$, 
which possibly should be constructed by the help of the primary operator 
$W_0 = \frac{1}{4}\, {\rm tr}\, \phi^4$ and with $\Psi_{\rm ext}$ being 
uniquely fixed by the requirement under vector supersymmetry, 
$Q_a \Psi_{\rm ext} = 0$.

\section{Dimensional reduction to four dimensions}

In order to get an idea how such an extension $S_{\rm ext}$ could be
constructed with the help of the primary operator 
$W_0 = \frac{1}{4}\,{\rm tr}\, \phi^4$, 
we look at the matter--independent part of the topologically half--twisted 
theory which, similarly to Ref. \cite{12}, is obtained by reducing to 
four dimensions the $Spin(7)$--invariant theory. To this end, 
for $1 \leq a,b \leq 4$, we group the components of $A_a$, $\psi_a$ and 
$\chi_{ab}$ into a vector isosinglet $A_A$, a vector isosinglet $\psi_A$ and 
a self--dual tensor isosinglet 
$\chi_{AB} = \frac{1}{2} \epsilon_{ABCD} \chi^{CD}$ ($A,B = 1,2,3,4$),
respectively, and for $5 \leq a,b \leq 8$, into a left--handed two-spinor 
isodoublet $G_\alpha^{\!~~i}$, a left--handed isodoublet
$\lambda_\alpha^{\!~~i}$ and a right--handed isodoublet
$\zeta_{\dot{\alpha}}^{\!~~i}$ ($i = 1,2$), respectively
(for the two--spinor conventions, see, Appendix A).
The index $i$ is raised and lowered as follows:
$\epsilon^{ij} \varphi_j = \varphi^i$ and 
$\varphi^i \epsilon_{ij} = \varphi_j$, where $\epsilon_{ij}$ is the 
invariant tensor of the group $SU(2)$, $\epsilon^{12} = \epsilon_{12} = 1$.

Then, for that action of the half--twisted theory one gets, according
to the prediction \cite{5}, 
\begin{align}
\label{3.1}
S_{\rm red} = S_0 + \int_E d^4x\, {\rm tr} \Bigr\{&
\hbox{$\frac{1}{2}$} D^A G^\alpha_{\!~~i} D_A G_\alpha^{\!~~i} +
\hbox{$\frac{1}{4}$} [ G^\alpha_{\!~~i}, G^\beta_{\!~~j} ]
[ G_\alpha^{\!~~i}, G_\beta^{\!~~j} ] -
2 [ G^\alpha_{\!~~i}, \bar{\phi} ] [ G_\alpha^{\!~~i}, \phi ]
\nonumber
\\
& + 2 \bar{\phi} \{ \lambda^\alpha_{\!~~i}, \lambda_\alpha^{\!~~i} \} +
2 \phi \{ \zeta^{\dot{\alpha} i}, \zeta_{\dot{\alpha} i} \} -
\hbox{$\frac{1}{2}$} \chi^{AB} (\sigma_{AB})^{\alpha\beta}
[ G_\alpha^{\!~~i}, \lambda_{\beta i} ]
\nonumber
\phantom{\frac{1}{2}}
\\
& - 2 i \zeta^{\dot{\alpha} i} (\sigma^A)_{\dot{\alpha} \beta}
D_A \lambda^\beta_{\!~~i} +
2 i \zeta^{\dot{\alpha} i} (\sigma^A)_{\dot{\alpha} \beta}
[ G^\beta_{\!~~i}, \psi_A ] +
2 \eta [ G^\alpha_{\!~~i}, \lambda_\alpha^{\!~~i} ] \Bigr\},
\end{align}
where the matter--independent part $S_0$ is just the Donaldson--Witten 
action \cite{4},
\begin{align}
\label{3.2}
S_0 = \int_E d^4x\, {\rm tr} \Bigr\{&
\hbox{$\frac{1}{8}$} \Theta^{ABCD} F_{AB} F_{CD} -
2 D^A \bar{\phi} D_A \phi - 2 \chi^{AB} D_A \psi_B + 2 \eta D^A \psi_A
\nonumber
\\
& + 2 \bar{\phi} \{ \psi^A, \psi_A \} +
\hbox{$\frac{1}{2}$} \phi \{ \chi^{AB}, \chi_{AB} \} +
2 \phi \{ \eta, \eta \} - 2 [ \phi, \bar{\phi} ]^2 \Bigr\}.
\end{align}
Hence, the result of compactifying the $Spin(7)$--invariant theory (\ref{2.1}) 
to four dimensions gives the Donaldson--Witten theory with matter in the 
adjoint representation. On the other hand, Donaldson--Witten theory with 
matter, after some rearrangements of the spinor fields, so that $S_{\rm red}$ 
becomes real, gets unified in eight dimensions (the resulting theory is very 
similar to the non--Abelian version of the Seiberg--Witten monopole theory). 

The (unnormalized) projector
\begin{equation}
\label{3.3}
\Theta_{ABCD} = \delta_{AC} \delta_{BD} - \delta_{AD} \delta_{BC} +
\epsilon_{ABCD},
\qquad
\hbox{$\frac{1}{4}$} \Theta_{ABEF} \Theta_{CD}^{~~~~EF} = \Theta_{ABCD},
\end{equation}
projects any antisymmetric second rank tensor onto its self--dual part
$\mathbf{3}$ according to the decomposition
$\mathbf{6} = \mathbf{3} \oplus \mathbf{3}$ of the adjoint representation
of $SO(4) \sim Spin(3) \otimes Spin(3)$. It obeys the relations
\begin{align}
\label{3.4}
\hbox{$\frac{1}{2}$} ( \Theta_{ABEG} \Theta_{CDF}^{~~~~~~G} -
\Theta_{ABFG} \Theta_{CDE}^{~~~~~~G} ) &=
\Theta_{EFAC} \delta_{BD} - \Theta_{EFAD} \delta_{BC}
\nonumber
\\
&\quad - \Theta_{EFBC} \delta_{AD} + \Theta_{EFBD} \delta_{AC}
\nonumber
\\
&= \Theta_{ABCE} \delta_{DF} - \Theta_{ABDE} \delta_{CF}
\nonumber
\\
&\quad - \Theta_{ABCF} \delta_{DE} + \Theta_{ABDF} \delta_{CE},
\nonumber
\\
\hbox{$\frac{1}{2}$} ( \Theta_{ABEG} \Theta_{CDF}^{~~~~~~G} +
\Theta_{ABFG} \Theta_{CDE}^{~~~~~~G} ) &= \Theta_{ABCD} \delta_{EF},
\end{align}
with $\epsilon_{ABCD}$ being the Levi--Civita tensor in four dimensions.

Furthermore, for the dimensionally reduced transformation rules,
generated by $Q$, $Q_A$ and $Q_{AB} = \frac{1}{2} \epsilon_{ABCD} Q^{CD}$, 
from (\ref{2.4})--(\ref{2.6}) one gets
\begin{align}
\label{3.5}
&Q A_A = \psi_A,
\nonumber
\\
&Q \psi_A = D_A \phi,
\nonumber
\\
&Q \phi = 0,
\nonumber
\\
&Q \bar{\phi} = \eta,
\nonumber
\\
&Q \eta = [ \bar{\phi}, \phi ],
\nonumber
\\
&Q \chi_{AB} = \hbox{$\frac{1}{4}$} \Theta_{ABCD} F^{CD} +
\hbox{$\frac{1}{4}$} (\sigma_{AB})^{\alpha\beta}
[ G_\alpha^{\!~~i}, G_{\beta i} ],
\nonumber
\\
&Q G_\alpha^{\!~~i} = \lambda_\alpha^{\!~~i},
\nonumber
\\
&Q \lambda_\alpha^{\!~~i} = [ G_\alpha^{\!~~i}, \phi ],
\nonumber
\\
&Q \zeta_{\dot{\alpha}}^{\!~~i} = \hbox{$\frac{1}{2}$}
i (\sigma^A)_{\beta \dot{\alpha}} D_A G^{\beta i},
\\
\intertext{}
\label{3.6}
&Q_A A_B = \delta_{AB} \eta + \chi_{AB},
\nonumber
\\
&Q_A \psi_B = F_{AB} - \hbox{$\frac{1}{4}$} \Theta_{ABCD} F^{CD} +
\delta_{AB} [ \phi, \bar{\phi} ] -
\hbox{$\frac{1}{4}$} (\sigma_{AB})^{\alpha\beta}
[ G_\alpha^{\!~~i}, G_{\beta i} ],
\nonumber
\\
&Q_A \phi = \psi_A,
\nonumber
\\
&Q_A \bar{\phi} = 0,
\nonumber
\\
&Q_A \eta = D_A \bar{\phi},
\nonumber
\\
&Q_A \chi_{CD} = \Theta_{ABCD} D^B \bar{\phi},
\nonumber
\\
&Q_A G_\alpha^{\!~~i} = i (\sigma_A)_{\alpha \dot{\beta}}
\zeta^{\dot{\beta} i},
\nonumber
\\
&Q_A \lambda_\alpha^{\!~~i} = \hbox{$\frac{1}{2}$} D_A G_\alpha^{\!~~i} -
\hbox{$\frac{1}{2}$} (\sigma_{AB})_{\alpha\beta} D^B G^{\beta i},
\nonumber
\\
&Q_A \zeta_{\dot{\alpha}}^{\!~~i} = i (\sigma_A)_{\beta \dot{\alpha}}
[ G^{\beta i}, \bar{\phi} ],
\\
\intertext{and}
\label{3.7}
&Q_{AB} A_C = - \Theta_{ABCD} \psi^D,
\nonumber
\\
&Q_{AB} \psi_C = \Theta_{ABCD} D^D \phi,
\nonumber
\\
&Q_{AB} \phi = 0,
\nonumber
\\
&Q_{AB} \bar{\phi} = \chi_{AB},
\nonumber
\\
&Q_{AB} \eta = - \hbox{$\frac{1}{4}$} \Theta_{ABCD} F^{CD} -
\hbox{$\frac{1}{4}$} (\sigma_{AB})^{\alpha\beta}
[ G_\alpha^{\!~~i}, G_{\beta i} ],
\nonumber
\\
&Q_{AB} \chi_{CD} = \hbox{$\frac{1}{4}$} \Theta_{ABEG} \Theta_{CDFG} F^{EF} +
\Theta_{ABCD} [ \bar{\phi}, \phi ] +
\hbox{$\frac{1}{4}$} (\sigma_{AB})^\alpha_{\!~~\gamma}
(\sigma_{CD})^{\beta\gamma} [ G_\alpha^{\!~~i}, G_{\beta i} ],
\nonumber
\\
&Q_{AB} G_\alpha^{\!~~i} = (\sigma_{AB})_{\alpha\beta} \lambda^{\beta i},
\nonumber
\\
&Q_{AB} \lambda_\alpha^{\!~~i} = - (\sigma_{AB})_{\alpha\beta}
[ G^{\beta i}, \phi ],
\nonumber
\\
&Q_{AB} \zeta_{\dot{\alpha}}^{\!~~i} = \hbox{$\frac{1}{4}$}
i (\sigma^C)_{\beta \dot{\alpha}} \Theta_{ABCD} D^D G^{\beta i}.
\end{align}
Notice that by introducing an auxiliary field 
$B_{AB} = \frac{1}{2} \epsilon_{ABCD} B^{CD}$ the matter--independent part 
of (\ref{3.5})--(\ref{3.7}) can be closed {\it off--shell} (see, Appendix B).

The crucial point is that, on--shell, the matter--independent part $S_0$ can
be obtained from the operator $\hat{W}_0 = \frac{1}{2}\, {\rm tr}\, \phi^2$ 
as follows,
\begin{equation}
\label{3.8}
S_0 \doteq S_{\rm top}^{4D} - 
\hbox{$\frac{1}{4!}$} \epsilon^{ABCD} Q_A Q_B Q_C Q_D
\int_E d^4x\, \hbox{$\frac{1}{2}$}\, {\rm tr}\, \phi^2 \doteq Q \Psi_0,
\end{equation}
where
\begin{equation}
\label{3.9}
S_{\rm top}^{4D} = \int_E d^4x\, {\rm tr} \Bigr\{
\hbox{$\frac{1}{8}$} \epsilon^{ABCD} F_{AB} F_{CD} \Bigr\}
\end{equation}
is the four--dimensional Pontryagin invariant. Moreover, on--shell,
the gauge fermion $\Psi_0$ can be obtained from the prepotential
$V_0 = \hbox{$\frac{1}{2}$}\, {\rm tr}\, \bar{\phi}^2$ according to
\begin{equation}
\label{3.10}
\Psi_0 \doteq \hbox{$\frac{1}{4!}$} Q_A^{\!~~B} Q_B^{\!~~C} Q_C^{\!~~A}
\int_E d^4x\, \hbox{$\frac{1}{2}$}\, {\rm tr}\, \bar{\phi}^2.
\end{equation}

On the other hand, the topological observables of the Donaldson--Witten
theory can also be obtained from $\hat{W}_0$ by means of the following ladder 
of $k$--forms $\hat{W}_k$ ($0 \leq k \leq 4$) with ghost number 
$4 - k$ \cite{4},
\begin{align}
\label{3.11}
&\hat{W}_0 = {\rm tr}\, ( \hbox{$\frac{1}{2}$} \phi^2 ),
\nonumber
\\
&\hat{W}_1 = {\rm tr}\, ( \psi \phi ),
\nonumber
\\
&\hat{W}_2 = {\rm tr}\, ( F \phi + \hbox{$\frac{1}{2}$} \psi \wedge \psi ),
\nonumber
\\
&\hat{W}_3 = {\rm tr}\, ( F \wedge \psi + \psi \wedge \psi ),
\nonumber
\\
&\hat{W}_4 = {\rm tr}\, ( \hbox{$\frac{1}{2}$} F \wedge F ).
\end{align}
Hence, because (\ref{2.9}) is just the eight--dimensional analogue of
(\ref{3.11}) the cohomological extension $S_{\rm ext}$ of the action 
(\ref{2.1}) should be precisely the eight--dimensional extension of 
(\ref{3.8})--(\ref{3.10}).

\section{Cohomological extension $S_{\rm ext}$ in the Landau type gauge}

As anticipated, for the cohomological extension $S_{\rm ext}$ we are 
looking for we make the following ansatz
\begin{equation}
\label{4.1}
S_{\rm ext} \doteq S_{\rm top}^{8D} -
\hbox{$\frac{1}{8!}$} \epsilon^{abcdefgh} Q_a Q_b Q_c Q_d Q_e Q_f Q_g Q_h
\int_E d^8x\, \hbox{$\frac{1}{4}$}\, {\rm tr}\, \phi^4 \doteq
Q \Psi_{\rm ext},
\end{equation}
with
\begin{equation}
\label{4.2}
\Psi_{\rm ext} \doteq \hbox{$\frac{1}{8!}$} 
Q_a^{\!~~b} Q_b^{\!~~c} Q_c^{\!~~d}
Q_d^{\!~~e} Q_e^{\!~~f} Q_f^{\!~~g} Q_g^{\!~~a}
\int d^8x\, \hbox{$\frac{1}{4}$}\, {\rm tr}\, \bar{\phi}^4,
\end{equation}
where
\begin{equation}
\label{4.3}
S_{\rm top}^{8D} = \int_E d^8x\, {\rm tr} \Bigr\{
\hbox{$\frac{1}{64}$} \epsilon^{abcdefgh} F_{ab} F_{cd} F_{ef} F_{gh} \Bigr\}
\end{equation}
is the eight--dimensional analogue of the Pontryagin invariant.

Let us briefly comment on the possibilities to determine $\Psi_{\rm ext}$ 
either from (\ref{4.1}) or from (\ref{4.2}). The most favorable way seems 
to be to construct $S_{\rm ext}$ by means of $Q$ and $Q_{ab}$ 
(and not by means of $Q_a$) because then $\Psi_{\rm ext}$ can be obtained 
directly from the prepotential $V_0 = \frac{1}{4}\, {\rm tr}\, \bar{\phi}^4$. 
However, despite the fact that the transformation rules 
(\ref{2.4})--(\ref{2.6}) look very similar to those of the matter--independent 
part of (\ref{3.5})--(\ref{3.7}), it is impossible to close the latter
{\it off--shell} with a finite number of auxiliary fields. Apart
from the general arguments given in \cite{16}, this may be traced
back to the fact that in the former case there are fewer algebraic
identities among the $Spin(7)$ projection operator $\Theta_{abcd}$
than for the $Spin(3)$ projection operator $\Theta_{ABCD}$ in the latter case
(c.f., Eq. (\ref{2.2}) with Eq. (\ref{3.4})).

More precisely, we were able to find an off--shell realization only for the 
subset $Q$, $Q_{ab}$ (see, Appendix C), but not for the subset 
$Q$, $Q_a$. For that reason, we do not further pursue the possibility 
to determine $\Psi_{\rm ext}$ by means of (\ref{4.2}). Hence, for our purpose 
only the relationship (\ref{4.1}) is really eligible --- after recasting it 
into an $Q$--exact form. However, proceeding in that less ideal way one 
is confronted with the tricky problem to verify the $Q$--exactness of 
(\ref{4.1}) and, therefore, the exposition of $\Psi_{\rm ext}$ becomes 
very complex (see, Appendix C).

Obviously, when evaluating (\ref{4.1}) in the Feynman type gauge one
gets a huge number of terms belonging to the non--minimal sector which, 
however, are of no particular interest. Therefore, focusing only on terms 
belonging to minimal sector, we shall restrict ourselves to the Landau 
type gauge. In that gauge the action (\ref{2.1}) simplifies as follows:
\begin{equation}
\label{4.4}
S^\prime = Q \Psi^\prime,
\qquad
\Psi^\prime = \int_E d^8x\, {\rm tr} \Bigr\{
\chi^{ab} F_{ab} - 2 \psi^a D_a \bar{\phi} \Bigr\},
\end{equation}
where the first term of $\Psi^\prime$ enforces the localization into the 
moduli space whereas the second term ensures that pure gauge degrees of 
freedom are projected out. One easily verifies that $\Psi^\prime$ is 
invariant under the vector supersymmetry $Q_a^\prime$ 
(in the Landau type gauge)
\begin{alignat}{2}
\label{4.5}
&Q_a^\prime A_b = 0,
&\qquad
&Q_a^\prime \bar{\phi} = 0,
\nonumber
\\
&Q_a^\prime \phi = \psi_a,
&\qquad
&Q_a^\prime \eta = D_a \bar{\phi},
\nonumber
\\
&Q_a^\prime \psi_b = F_{ab},
&\qquad
&Q_a^\prime \chi_{cd} = \Theta_{abcd} D^b \bar{\phi}.
\end{alignat}
For the cohomological extension $S_{\rm ext}^\prime$ of (\ref{4.4}), 
whose computation is postponed to Appendix C, one obtains 
\begin{equation}
\label{4.6}
S_{\rm ext}^\prime = Q \Psi_{\rm ext}^\prime,
\qquad
\Psi_{\rm ext}^\prime = \alpha \int_E d^8x\, {\rm tr}_{\rm (s)} \Bigr\{
\Phi^{[abcd} \chi^{ef]} F_{ab} F_{cd} F_{ef} -
12 \psi^e \Phi_{[abcd} D_{e]} \bar{\phi} F^{ab} F^{cd} \Bigr\},
\end{equation}
$\alpha$ being an arbitrary constant. $\Psi_{\rm ext}^\prime$ is invariant 
under the vector supersymmetry (\ref{4.5}) as well.
\footnote{The square bracket antisymmetrization is iteratively defined as
\begin{align*}
&[ ab ] = ab - ba, \quad
[ abc ] = a [ bc ] + b [ ca ] + c [ ab ], \quad
[ abcd ] = a [ bcd ] - b [ cda ] + c [ dab ] - d [ abc ],
\quad{\rm etc.}
\end{align*}}
In order to check this crucial property one needs the following identities:
\begin{align}
\label{4.7}
\hbox{$\frac{1}{2}$} \epsilon^{abcdefgh} \Phi_{mngh} &=
\Phi^{abcd} \delta^e_{[m} \delta^f_{n]} -
\Phi^{bcd[e} \delta^{f]}_{[m} \delta^a_{n]} +
\hbox{$\frac{1}{2}$} \Phi^{cd[ef} \delta^{a]}_{[m} \delta^b_{n]}
\nonumber
\\
&\quad - \hbox{$\frac{1}{6}$} \Phi^{d[efa} \delta^{b]}_{[m} \delta^c_{n]} +
\hbox{$\frac{1}{24}$} \Phi^{[efab} \delta^{c]}_{[m} \delta^d_{n]} =
\hbox{$\frac{1}{48}$} \Phi^{[abcd} \delta^e_{[m} \delta^{f]}_{n]},
\nonumber
\\
\hbox{$\frac{1}{6}$} \epsilon^{abcdefgh} \Phi_{mfgh} &=
\hbox{$\frac{1}{24}$} \Phi^{[abcd} \delta^{e]}_m,
\nonumber
\\
\hbox{$\frac{1}{24}$} \epsilon^{abcdefgh} \Phi_{efgh} &=
\Phi^{abcd},
\end{align}
where the last one expresses the self--duality of $\Phi_{abcd}$.

It is amusing to see that, in the Landau type gauge, the extension of
$\Psi^\prime$ into $\Psi_{\rm ext}^\prime$ is rather simple. Namely,
it obtains either by performing in (\ref{4.4}) the replacements
\begin{gather*}
\Phi_{abef} F^{ab} \rightarrow \Phi_{abef} F^{ab} +
\alpha\, \Phi^{abcd} F_{[ab} F_{cd} F_{ef]},
\\
D^e \psi_e \rightarrow D^e \psi_e +
6 \alpha\, \Phi^{[abcd} D^{e]} (
\psi_e F_{ab} F_{cd} + F_{ab} \psi_e F_{cd} + F_{ab} F_{cd} \psi_e ),
\end{gather*}
or, equivalently, by changing (formally) in the same manner the self--duality
gauge condition $\Phi_{abef} F^{ab} = 0$ and the ghost gauge condition
$D^e \psi_e = 0$.

Summarizing, we have shown that the $Spin(7)$--invariant super Yang--Mills
theory, which relies on the existence of the Cayley invariant, permits the 
construction of a cohomological extension by the help of the operator 
$W_0 = \frac{1}{4}\,{\rm tr}\, \phi^4$, which relies on the existence of the
eight--dimensional analogue of the Pontryagin invariant.

With regard to this, a couple of interesting questions is left still open 
deserving a further study. So far, the $Spin(7)$--invariant theory was 
considered in flat space only. But, according to Berger's classification 
\cite{17} a metric with $Spin(7)$ holonomy on the simply connected 
eight--dimensional Riemannian manifold $\cal{M}$ with Euclidean signature 
admits a covariantly--constant spinor $\zeta$. If such $\zeta$ exists, the
metric is automatically Ricci--flat. In addition, such metric has the 
$Spin(7)$--invariant closed Cayley four--form $\Phi$ (for a given choice of 
the orientation of $\cal{M}$). Conversely, if $\Phi$ with respect
to a metric on $\cal{M}$ is closed, then the metric has $Spin(7)$ holonomy. 
Hence, the action (\ref{4.4}) and its extension (\ref{4.6}) can be considered 
on a curved manifold $\cal{M}$ with $Spin(7)$ holonomy. 

As is widely believed, super Yang--Mills theory in eight dimensions may arise 
as low--energy effective world volume theory on a Euclidean 7--brane in 
Type IIB string theory \cite{9}. Thus, as it was pointed out in \cite{6},
the $Spin(7)$--invariant theory in curved space can be considered as a 
theory being obtained by wrapping an Euclidean 7--brane of Type IIB string 
theory around a manifold with $Spin(7)$ holonomy. However, such a theory is 
not renormalizable and, therefore, extra degrees of freedom are needed at high
energies. It seems to be natural to assume that such extra degrees of
freedom are given by the counterterms arising from string theory after
compactification to eight dimensions. In another context, in \cite{11}
some arguments were given that all the string--corrected eight--dimensional
counterterms are still $Q$--exact. So, one may ask whether in
the Landau type gauge the cohomological extension (\ref{4.6}) in
curved space differs from the string--corrected one--loop counterterm 
(identifying $\alpha$ with the string tension) only by replacing the 
Levi--Civity tensor $\epsilon_{abcdefgh}$ through a certain $SO(8)$--invariant 
tensor $t_{abcdefgh}(\phi^2)$ involving the operator 
$W_0 = \frac{1}{2}\, {\rm tr}\, \phi^2$ (see, e.g., \cite{18}). Another
question is whether, in a similar way, a cohomological extension can
be constructed from the operator $W_0 = \frac{1}{6}\, {\rm tr}\, \phi^6$, 
too, which should be related to the string--corrected two--loop counterterms, 
and so on.

\begin{flushleft}
{\large{\bf Appendix A}}
\end{flushleft}
\medskip
The Euclidean two--spinor conventions adopted in this paper are similar 
to those of Ref. \cite{19}, Appendix E. The numerically invariant tensors
$(\sigma_A)^{\alpha \dot{\beta}}$ and $(\sigma_A)_{\dot{\alpha} \beta}$
are the Clebsch--Cordon coefficients relating the
$( \frac{1}{2}, \frac{1}{2} )$ representation of $SL(2,\mathbb{C})$
to the vector representation of $SO(4)$,
\begin{alignat*}{2}
&(\sigma_A)^{\dot{\alpha} \beta} =
( -i \sigma_1, -i \sigma_2, -i \sigma_3, I_2 ),
&\qquad
&(\sigma_A)_{\dot{\alpha} \beta} :=
(\sigma_A)^{\dot{\gamma} \delta}
\epsilon_{\dot{\gamma}\dot{\alpha}} \epsilon_{\delta\beta} =
(\sigma_A^*)^{\dot{\alpha} \beta},
\\
&(\sigma_A)_{\alpha \dot{\beta}} =
( i \sigma_1, i \sigma_2, i \sigma_3, I_2 ),
&\qquad
&(\sigma_A)^{\alpha \dot{\beta}} :=
\epsilon^{\alpha\gamma} \epsilon^{\dot{\beta}\dot{\delta}}
(\sigma_A)_{\gamma \dot{\delta}} =
(\sigma_A^*)_{\alpha \dot{\beta}},
\end{alignat*}
$(\sigma_A)_{\dot{\alpha} \beta}$ and $(\sigma_A)^{\alpha \dot{\beta}}$
being the corresponding complex conjugate coefficients. Thereby, $\sigma_1$,
$\sigma_2$, $\sigma_3$ are the Pauli matrices,
which satisfy the Clifford algebra
\begin{align*}
&(\sigma_A)^{\alpha \dot{\gamma}} (\sigma_B)_{\dot{\gamma} \beta} +
(\sigma_B)^{\alpha \dot{\gamma}} (\sigma_A)_{\dot{\gamma} \beta} =
2 \delta_{AB} \delta^\alpha_{\!~~\beta},
\\
&(\sigma_A)_{\dot{\alpha} \gamma} (\sigma_B)^{\gamma \dot{\beta}} +
(\sigma_B)_{\dot{\alpha} \gamma} (\sigma_A)^{\gamma \dot{\beta}} =
2 \delta_{AB} \delta^{\dot{\alpha}}_{\!~~\dot{\beta}},
\end{align*}
and, in addition, the completeness relations,
\begin{alignat*}{2}
&(\sigma_A)_{\dot{\alpha} \beta} (\sigma^B)^{\beta \dot{\alpha}} =
2 \delta_A^{\!~~B},
&\qquad
&(\sigma_A)_{\dot{\alpha} \beta} (\sigma^A)^{\gamma \dot{\delta}} =
2 \delta_{\dot{\alpha}}^{\!~~\dot{\delta}} \delta_\beta^{\!~~\gamma},
\\
&(\sigma_A)_{\dot{\alpha} \beta} (\sigma^A)_{\dot{\gamma} \delta} =
2 \epsilon_{\dot{\alpha}\dot{\gamma}} \epsilon_{\beta\delta},
&\qquad
&(\sigma_A)^{\alpha \dot{\beta}} (\sigma^A)^{\gamma \dot{\delta}} =
2 \epsilon^{\alpha\gamma} \epsilon^{\dot{\beta}\dot{\delta}}.
\end{alignat*}
The spinor index $\alpha$ (analogously $\dot{\alpha}$) is raised and
lowered by $\epsilon^{\alpha\gamma} \varphi_\gamma^{\!~~\beta} =
\varphi^{\alpha\beta}$ and $\varphi_\alpha^{\!~~\gamma}
\epsilon_{\gamma\beta} = \varphi_{\alpha\beta}$,
where $\epsilon_{\alpha\beta}$ (analogously
$\epsilon_{\dot{\alpha}\dot{\beta}}$) is the invariant tensor of the
group $SU(2)$, $\epsilon_{12} = \epsilon^{12} =
\epsilon_{\dot{1}\dot{2}} = \epsilon^{\dot{1}\dot{2}} = 1$.

The selfdual and anti--selfdual $SO(4)$ generators,
$(\sigma_{AB})_{\alpha\beta}$ and $(\sigma_{AB})_{\dot{\alpha}\dot{\beta}}$,
and the various Clebsch--Gordon coefficients are related by the properties
\begin{align*}
&(\sigma_A)^{\alpha \dot{\gamma}} (\sigma_B)_{\dot{\gamma}}^{\!~~\beta} =
(\sigma_{AB})^{\alpha\beta} -
\delta_{AB} \epsilon^{\alpha\beta},
\\
&(\sigma_C)^{\alpha \dot{\gamma}}
(\sigma_{AB})_{\dot{\gamma}}^{\!~~\dot{\beta}} =
( \delta_{AC} \delta_{BD} - \delta_{BC} \delta_{AD} -
\epsilon_{ABCD} ) (\sigma^D)^{\alpha \dot{\beta}},
\\
&(\sigma_{CD})^{\dot{\alpha}\dot{\gamma}}
(\sigma_{AB})_{\dot{\gamma}}^{\!~~\dot{\beta}} =
( \delta_{AC} \delta_{BD} - \delta_{BC} \delta_{AD} -
\epsilon_{ABCD} ) \epsilon^{\dot{\alpha}\dot{\beta}} -
\delta_{[C[A} (\sigma_{B]D]})^{\dot{\alpha}\dot{\beta}},
\\
&(\sigma_A)_{\dot{\alpha} \gamma}
(\sigma_B)^\gamma_{\!~~\dot{\beta}} =
(\sigma_{AB})_{\dot{\alpha}\dot{\beta}} +
\delta_{AB} \epsilon_{\dot{\alpha}\dot{\beta}},
\\
&(\sigma_C)_{\dot{\alpha} \gamma}
(\sigma_{AB})^\gamma_{\!~~\beta} =
( \delta_{AC} \delta_{BD} - \delta_{BC} \delta_{AD} +
\epsilon_{ABCD} ) (\sigma^D)_{\dot{\alpha} \beta},
\\
&(\sigma_{CD})_{\alpha\gamma}
(\sigma_{AB})^\gamma_{\!~~\beta} =
( \delta_{AC} \delta_{BD} - \delta_{BC} \delta_{AD} +
\epsilon_{ABCD} ) \epsilon_{\alpha\beta} -
\delta_{[C[A} (\sigma_{B]D]})_{\alpha\beta}.
\end{align*}

Finally, some often used identities are
\begin{alignat*}{2}
&(\sigma_{AB})_{\dot{\alpha}\dot{\beta}}
(\sigma^B)_{\dot{\gamma} \delta} =
- 2 (\sigma_A)_{\dot{\alpha} \delta} \epsilon_{\dot{\beta}\dot{\gamma}} -
\epsilon_{\dot{\alpha}\dot{\beta}} (\sigma_A)_{\dot{\gamma} \delta},
&\qquad
&(\sigma^{AB})_{\dot{\alpha}\dot{\beta}}
(\sigma_{AB})_{\dot{\gamma}\dot{\delta}} =
8 \epsilon_{\dot{\alpha}\dot{\gamma}} \epsilon_{\dot{\beta}\dot{\delta}} -
4 \epsilon_{\dot{\alpha}\dot{\beta}} \epsilon_{\dot{\gamma}\dot{\delta}},
\\
&(\sigma_{AB})_{\alpha\beta}
(\sigma^B)_{\gamma \dot{\delta}} =
2 (\sigma_A)_{\alpha \dot{\delta}} \epsilon_{\beta\gamma} +
\epsilon_{\alpha\beta} (\sigma_A)_{\gamma \dot{\delta}},
&\qquad
&(\sigma^{AB})_{\alpha\beta}
(\sigma_{AB})_{\gamma\delta} =
8 \epsilon_{\alpha\gamma} \epsilon_{\beta\delta} -
4 \epsilon_{\alpha\beta} \epsilon_{\gamma\delta}.
\end{alignat*}

\bigskip
\begin{flushleft}
{\large{\bf Appendix B}}
\end{flushleft}
\medskip
In this Appendix we give the {\it off--shell} transformation rules of the
matter--independent part of (\ref{3.5})--(\ref{3.7}),
\begin{align}
&Q A_A = \psi_A,
\nonumber
\\
&Q \psi_A = D_A \phi,
\nonumber
\\
&Q \phi = 0,
\nonumber
\\
&Q \bar{\phi} = \eta,
\nonumber
\\
&Q \eta = [ \bar{\phi}, \phi ],
\nonumber
\\
&Q \chi_{AB} = B_{AB},
\nonumber
\\
&Q B_{AB} = [ \chi_{AB}, \phi ],
\tag{B.1}
\\
\intertext{}
&Q_A A_B = \delta_{AB} \eta + \chi_{AB}, \nonumber
\nonumber
\\
&Q_A \psi_B = F_{AB} - B_{AB} + \delta_{AB} [ \phi, \bar{\phi} ],
\nonumber
\\
&Q_A \phi = \psi_A,
\nonumber
\\
&Q_A \bar{\phi} = 0,
\nonumber
\\
&Q_A \eta = D_A \bar{\phi},
\nonumber
\\
&Q_A \chi_{CD} = \Theta_{ABCD} D^B \bar{\phi},
\nonumber
\\
&Q_A B_{CD} = D_A \chi_{CD} -
\Theta_{ABCD} ( [ \psi^B, \bar{\phi} ] + D^B \eta )
\tag{B.2}
\\
\intertext{} &Q_{AB} A_C = - \Theta_{ABCD} \psi^D, \nonumber
\\
&Q_{AB} \psi_C = \Theta_{ABCD} D^D \phi,
\nonumber
\\
&Q_{AB} \phi = 0,
\nonumber
\\
&Q_{AB} \bar{\phi} = \chi_{AB},
\nonumber
\\
&Q_{AB} \eta = - B_{AB},
\nonumber
\\
&Q_{AB} \chi_{CD} = - \Theta_{ABC}^{~~~~~~E}
( F_{DE} - \hbox{$\frac{1}{2}$} B_{DE} )
\nonumber
\\
&~\phantom{Q_{AB} \chi_{CD} =}
+ \Theta_{ABD}^{~~~~~~E} ( F_{CE} - \hbox{$\frac{1}{2}$} B_{CE} ) +
\Theta_{ABCD} [ \bar{\phi}, \phi ],
\nonumber
\\
&Q_{AB} B_{CD} = \Theta_{ABC}^{~~~~~~E}
( D_{[D} \psi_{E]} - \hbox{$\frac{1}{2}$} [ \chi_{DE}, \phi ] )
\nonumber
\\
&~\phantom{Q_{AB} B_{CD} =}
- \Theta_{ABD}^{~~~~~~E}
( D_{[C} \psi_{E]} - \hbox{$\frac{1}{2}$} [ \chi_{CE}, \phi ] ) -
\Theta_{ABCD} [ \eta, \phi ],
\tag{B.3}
\end{align}
where the operator $\Theta_{ABCD}$ was introduced in Eq. (\ref{3.3}).

\bigskip
\begin{flushleft}
{\large{\bf Appendix C}}
\end{flushleft}
\medskip
In this Appendix we prove the {\it off--shell} $Q$--exactness of the
cohomological extension (\ref{4.1}) and we show that, by choosing the Landau 
type gauge, one exactly reproduces the expression (\ref{4.6}) for the gauge
fermion $\Psi_{\rm ext}^\prime$.

As already emphasized in the Sect. 4, we are forced to start with the 
complicated expression
\begin{equation}
\tag{C.1}
S_{\rm ext} = S_{\rm top}^{8D} - \hbox{$\frac{1}{8!}$} \epsilon^{abcdefgh}
Q_a Q_b Q_c Q_d Q_e Q_f Q_g Q_h
\int_E d^8x\, \hbox{$\frac{1}{4}$}\, {\rm tr}\, \phi^4,
\end{equation}
where $S_{\rm top}^{8D}$ is the eight--dimensional topological invariant 
(\ref{4.3}). Namely, we found only the {\it off--shell} extension of the 
scalar and vector supersymmetry transformations (\ref{2.3}) and (\ref{2.4}), 
\begin{alignat}{2}
&Q A_a = \psi_a,
&\qquad
&Q \psi_a = D_a \phi,
\nonumber
\\
&Q \phi = 0,
&\qquad
&Q \bar{\phi} = \eta,
\nonumber
\\
&Q \eta = [ \bar{\phi}, \phi ],
&\qquad
&Q \chi_{ab} = B_{ab},
\nonumber
\\
&Q B_{ab} = [ \chi_{ab}, \phi ],
&&
\tag{C.2}
\\
\intertext{and}
&Q_a A_b = \delta_{ab} \eta + \chi_{ab},
&\qquad
&Q_a \psi_b = F_{ab} - B_{ab} + \delta_{ab} [ \phi, \bar{\phi} ],
\nonumber
\\
&Q_a \phi = \psi_a,
&\qquad
&Q_a \bar{\phi} = 0,
\nonumber
\\
&Q_a \eta = D_a \bar{\phi},
&\qquad
&Q_a \chi_{cd} = \Theta_{abcd} D^b \bar{\phi},
\nonumber
\\
&Q_a B_{cd} = D_a \chi_{cd} -
\Theta_{abcd} ( [ \psi^b, \bar{\phi} ] + D^b \eta ),
&&
\tag{C.3}
\end{alignat}
where $B_{ab} = \frac{1}{6} \Phi_{abcd} B^{cd}$ is the anti--field of
$\chi_{ab}$. One simply verifies that $Q$ and $Q_a$ satisfy the following
superalgebra {\it off--shell},
\begin{equation}
\tag{C.4}
\{ Q, Q \} = - 2 \delta_G(\phi),
\qquad
\{ Q, Q_a \} = \partial_a + \delta_G(A_a),
\qquad
\{ Q_a, Q_b \} = - 2 \delta_{ab} \delta_G(\bar{\phi}).
\end{equation}

In order to recast (C.1) into the $Q$--exact form
$S_{\rm ext} = Q \Psi_{\rm ext}$ it is convenient to exploit only the
algebraic relations (C.4) among $Q$ and $Q_a$, but not their explicit 
realizations (C.2) and (C.3). Otherwise, owing to the increasing number of 
terms which arise by evaluating (C.1), it is nearly impossible to determine
$\Psi_{\rm ext}$ explicitly.

To begin with, by repeated application of $Q_a$ on the primary operator
$W_0 = \frac{1}{4}\,{\rm tr}\, \phi^4$ we decompose (C.1) into a form where
the different operators $Q_a$ act only on the scalar field $\phi$. After  
straightforward calculations one obtains
\begin{align}
\tag{C.5}
S_{\rm ext} =\,& S_{\rm top}^{8D} -
\hbox{$\frac{1}{8!}$} \epsilon^{abcdefgh} \int_E d^8x\, {\rm tr}_{\rm (s)}
\Bigr\{
\psi_{abcdefgh} \phi^3 +
8 \psi_{abcdefg} \psi_h \phi^2 +
28 \psi_{abcdef} \psi_{gh} \phi^2
\nonumber
\\
& + 56 \psi_{abcdef} \psi_g \psi_h \phi +
56 \psi_{abcde} \psi_{fgh} \phi^2 +
168 \psi_{abcde} \psi_{fg} \psi_h \phi +
336 \psi_{abcde} \psi_f \psi_g \psi_h
\nonumber
\\
& + 35 \psi_{abcd} \psi_{efgh} \phi^2 +
280 \psi_{abcd} \psi_{efg} \psi_h \phi +
420 \psi_{abcd} \psi_{ef} \psi_{gh} \phi +
840 \psi_{abcd} \psi_{ef} \psi_g \psi_h
\nonumber
\phantom{\frac{1}{2}}
\\
& + 280 \psi_{abc} \psi_{def} \psi_{gh} \phi +
560 \psi_{abc} \psi_{def} \psi_g \psi_h +
1680 \psi_{abc} \psi_{de} \psi_{fg} \psi_h +
630 \psi_{ab} \psi_{cd} \psi_{ef} \psi_{gh} \Bigr\},
\nonumber
\end{align}
where
\begin{equation}
\tag{C.6}
\psi_a = Q_a \phi,
\qquad
\psi_{ab} = Q_a \psi_b,
\qquad
\psi_{abc} = Q_a \psi_{bc},\quad{\rm etc.},
\end{equation}
and where, for the sake simplicity, we have introduced a
{\it symmetrized} trace, ${\rm tr}_{\rm (s)}$. It is defined as follows: 
First, for every monomial $X_1 X_2 X_3 X_4$ in the integrand of (C.5) which,  
due to its origin from $\phi^4$, always consists of exactly 4 factors,
we consider all those graded permutations (with repetitions) of the various 
factors $X_i =(\phi, \psi_a, \psi_{ab}, \ldots$) which do not correspond 
to an antisymmetrization of their space indices; owing to the presence of the
Levi--Civita tensor in (C.5) we can ignore such permutations. After
that, we take the trace of the sum of all these permuted polynomials and
drop their largest common factor. 

As an illustration let us give some examples:  
\begin{align*}
&{\rm tr}_{\rm (s)} \Bigr\{ \psi_{abcdefg} \psi_{h} \phi^2 \Bigr\} =
{\rm tr} \Bigr\{
\psi_{abcdefg} ( \psi_h \phi^2 + \phi \psi_h \phi + \phi^2 \psi_h ) \Bigr\},
\\
&{\rm tr}_{\rm (s)} \Bigr\{ \psi_{abcd} \psi_{efgh} \phi^2 \Bigr\} =
{\rm tr} \Bigr\{
\psi_{abcd} ( \psi_{efgh} \phi^2 + \phi \psi_{efgh} \phi +
\phi^2 \psi_{efgh} ) \Bigr\},
\\
&{\rm tr}_{\rm (s)} \Bigr\{ \psi_{abcde} \psi_f \psi_g \psi_h \Bigr\} =
{\rm tr} \Bigr\{ \psi_{abcde} \psi_f \psi_g \psi_h \Bigr\}.
\end{align*}
In the first example, the 24 possible graded permutations of the monomial 
$\psi_{abcdefg} \psi_{h} \phi^2$, after taking the trace over their sum 
lead to only 3 different monomials with the common factor 8 which has to be 
dropped. In the same way one performs the symmetrized trace of the other 
examples. Notice, that in the second example one has to ignore the permutation 
of $\psi_{abcd}$ and $\psi_{efgh}$ in $\psi_{abcd} \psi_{efgh} \phi^2$ 
because it can be reversed through an antisymmetrization of their space 
indices. In the last example one even has to ignore all possible graded 
permutations.

With that definition of the symmetrized trace the various factors in front 
of the monomials in (C.5) agree precisely with the number of permutations 
which are necessary in order to recast the symmetrized trace of these 
polynomials into a fully antisymmetized form. For example, in order to 
recast $\psi_{abcd} \psi_{efgh} \phi^2$ into a totally antisymmetrized form 
one has to perform $8!/4!4! = 70$ permutations, thereby taking into account 
that, owing to the presence of the Levi--Civita tensor, only the completely 
antisymmetrized part of $\psi_{abcd}$ and $\psi_{efgh}$ appears in (C.5).
By taking the symmetrized trace of that polynomial this number is further
reduced to 35 since, under that trace, one can permute $\psi_{abcd}$ and
$\psi_{efgh}$ thereby dividing the total number of permutations by two.
Hence, one has to perform only 35 permutations in accordance with the
prefactor of that polynomial in (C.5).

As a next step, owing to the $Q$--exactness of $\psi_a$ we can split off from
each of the higher rank objects $\psi_{ab}$, $\psi_{abc}, \ldots$ in
(C.6) an $Q$--exact term by making use of the second relation
(C.4), as a result of which one gets the following decompositions,
\begin{align}
\tag{C.7}
&\psi_{ab} = - Q \lambda_{ab} + F_{ab},
\\
&\psi_{abc} = Q \lambda_{abc} - 3 D_a \lambda_{bc},
\nonumber
\\
&\psi_{abcd} = - Q \lambda_{abcd} + 4 D_a \lambda_{bcd} -
3 \{ \lambda_{ab}, \lambda_{cd} \},
\nonumber
\\
&\psi_{abcde} = Q \lambda_{abcde} - 5 D_a \lambda_{bcde} +
10 [ \lambda_{ab}, \lambda_{cde} ],
\nonumber
\\
&\psi_{abcdef} = - Q \lambda_{abcdef} + 6 D_a \lambda_{bcdef} +
10 [ \lambda_{abc}, \lambda_{def} ] - 15 \{ \lambda_{ab}, \lambda_{cdef} \},
\nonumber
\\
&\psi_{abcdefg} = Q \lambda_{abcdefg} -
7 D_a \lambda_{bcdefg} - 35 [ \lambda_{abc}, \lambda_{defg} ] +
21 [ \lambda_{ab}, \lambda_{cdefg} ],
\nonumber
\\
&\psi_{abcdefgh} = - Q \lambda_{abcdefgh} + 8 D_a \lambda_{bcdefgh} -
28 \{ \lambda_{ab}, \lambda_{cdefgh} \} +
56 [ \lambda_{abc}, \lambda_{defgh} ] -
35 \{ \lambda_{abcd}, \lambda_{efgh} \},
\nonumber
\end{align}
where
\begin{equation}
\tag{C.8}
\lambda_{ab} = Q_a A_b,
\qquad
\lambda_{abc} = Q_a \lambda_{bc},
\qquad
\lambda_{abcd} = Q_a \lambda_{bcd}, \quad {\rm etc.}
\end{equation}
Thereby, for the sake simplicity, we still have performed
some replacements on the right--hand side of (C.7). For example, let us
derive the decomposition of $\psi_{abc}$,
\begin{align*}
&\psi_a = Q_a \phi = Q A_a,
\\
&\psi_{ab} = Q_a \psi_b = Q_a ( Q A_b ) =
- Q ( Q_a A_b ) + F_{ab} = - Q \lambda_{ab} + F_{ab},
\\
&\psi_{abc} = Q_a \psi_{bc} = - Q_a ( Q \lambda_{bc} - F_{bc} ) =
Q ( Q_a \lambda_{bc} ) - D_a \lambda_{bc} -
D_{[c} ( Q_a A_{b]} ) \Rightarrow Q \lambda_{abc} - 3 D_a \lambda_{bc},
\end{align*}
where in the last relation we have replaced $D_{[c} ( Q_a A_{b]} ) =
D_c \lambda_{ab} - D_b \lambda_{ac}$ through $2 D_a \lambda_{bc}$ since only
the completely antisymmetric part of $\psi_{abc}$ enters into (C.5).
In exactly the same way one can derive iteratively all the other
decompositions in (C.7). Notice, that the various factors in front of
the different terms in (C.7) agree precisely with the number of
permutations which must be performed in order to fully antisymmetrize these
terms, thereby taking into account that after inserting the decompositions
(C.7) into (C.5) only the fully antisymmetric  part of
$\lambda_{ab}$, $\lambda_{abc}, \ldots$ contribute.

We are now faced with the difficult problem to rewrite (C.5) in an
$Q$--exact form. This is owing to the fact that after inserting the
decompositions (C.7) into (C.5) one gets a huge number of symmetrized terms
and, therefore, the computational effort in order to expose the gauge fermion
is considerably. By making use of the first relation (C.4), after a 
straightforward but lengthy algebraic computation for $\Psi_{\rm ext}$ one 
obtains
\begin{align}
\tag{C.9}
\Psi_{\rm ext} = \,& \hbox{$\frac{1}{8!}$} \epsilon^{abcdefgh}
\int_E d^8x\, {\rm tr}_{\rm (s)}  \Bigr\{
\lambda_{abcdefgh} \phi^3 -
8 \lambda_{abcdefg} \psi_h \phi^2 -
28 \lambda_{abcdef} ( Q \lambda_{gh} - F_{gh} ) \phi^2
\nonumber
\\
& + 56 \lambda_{abcdef} \psi_g \psi_h \phi -
56 \lambda_{abcde} ( Q \lambda_{fgh} - 3 D_f \lambda_{gh} ) \phi^2 -
336 \lambda_{abcde} \psi_f \psi_g \psi_h
\nonumber
\\
& + 168 \lambda_{abcde} \psi_f ( Q \lambda_{gh} - F_{gh} ) \phi -
35 \lambda_{abcd} ( Q \lambda_{efgh} - 8 D_e \lambda_{fgh} +
6 \{ \lambda_{ef}, \lambda_{gh} \} ) \phi^2
\nonumber
\phantom{\frac{1}{2}}
\\
& + 280 \lambda_{abcd} \psi_e ( Q \lambda_{fgh} - 3 D_f \lambda_{gh} ) \phi +
420 \lambda_{abcd} ( Q \lambda_{ef} - F_{ef} )( Q \lambda_{gh} - F_{gh} ) \phi
\nonumber
\\
& - 840 \lambda_{abcd} \psi_e \psi_f ( Q \lambda_{gh} - F_{gh} ) +
280 \lambda_{abc} ( Q \lambda_{def} - 6 D_d \lambda_{ef} )
( Q \lambda_{gh} - F_{gh} ) \phi
\nonumber
\phantom{\frac{1}{2}}
\\
& - 560 \lambda_{abc} \psi_d \psi_e ( Q \lambda_{fgh} - 6 D_f \lambda_{gh} ) -
1680 \lambda_{abc} \psi_d ( Q \lambda_{ef} - F_{ef} )
( Q \lambda_{gh} - F_{gh} )
\nonumber
\\
& + 280 \lambda_{abc} [ \lambda_{def}, \lambda_{gh} ] \phi^2 +
840 \lambda_{abc} \psi_d \{ \lambda_{ef}, \lambda_{gh} \} \phi -
420 \lambda_{ab} \{ \lambda_{cd}, \lambda_{ef} \} F_{gh} \phi
\nonumber
\phantom{\frac{1}{2}}
\\
& + 315 \lambda_{ab} \{ \lambda_{cd}, \lambda_{ef} \} Q \lambda_{gh} \phi -
630 \lambda_{ab} Q \lambda_{cd} Q \lambda_{ef} Q \lambda_{gh} -
560 \lambda_{abc} \psi_d D_e \lambda_{fgh} \phi
\nonumber
\\
& + 1680 \lambda_{ab} \psi_c D_d \lambda_{ef} Q \lambda_{gh} -
840 \lambda_{ab} \psi_c \psi_d \{ \lambda_{ef}, \lambda_{gh} \} +
840 \lambda_{ab} Q \lambda_{cd} Q \lambda_{ef} F_{gh}
\nonumber
\phantom{\frac{1}{2}}
\\
& + 1680 \lambda_{ab} D_g \lambda_{cd} D_h \lambda_{ef} \phi -
1260 \lambda_{ab} Q \lambda_{cd} F_{ef} F_{gh} -
2520 \lambda_{ab} \psi_c D_d \lambda_{ef} F_{gh}
\nonumber
\\
& + 2520 \lambda_{ab} F_{cd} F_{ef} F_{gh} \Bigr\},
\nonumber
\end{align}
which is the eight--dimensional extension of the gauge fermion $\Psi_0$ of
the Donaldson--Witten theory.

Here, a remark is in order. In (C.5) the symmetrized trace was
introduced for monomials consisting of exactly 4 factors $X_i$. 
After substituting for $\psi_{ab}$, $\psi_{abc}, \ldots$
the decompositions (C.7) we introduce, besides the field strength
$F_{ab}$, also the covariant derivative $D_a$ and some graded commutators of
the higher rank objects $\lambda_{ab}$, $\lambda_{abc}, \ldots$
as well as their $Q$--transforms. Thus, in order not to spoil the definition
of the symmetrized trace, one has to view these new objects, in particular
{\it all the graded commutators} of $\lambda_{ab}$, $\lambda_{abc}, \ldots$,
as single factors, i.e., new objects like $X_i$! Let us still give some 
example of the correct use of the symmetrized trace in (C.9),
\begin{align*}
&{\rm tr}_{\rm (s)} \Bigr\{ 
[ \lambda_{ab}, \lambda_{cde} ]
\lambda_{fgh} \phi^2 \Bigr\} = {\rm tr} \Bigr\{
[ \lambda_{ab}, \lambda_{cde} ] (
\lambda_{fgh} \phi^2 + \phi \lambda_{fgh} \phi +
\phi^2 \lambda_{fgh} ) \Bigr\},
\\
&{\rm tr}_{\rm (s)} \Bigr\{ 
\lambda_{abcd} \{ \lambda_{ef}, \lambda_{gh} \}
\phi^2 \Bigr\} = {\rm tr} \Bigr\{ 
\lambda_{abcd} \Bigr[ \{ \lambda_{ef}, \lambda_{gh} \} \phi^2 +
\phi \{ \lambda_{ef}, \lambda_{gh} \} \phi +
\phi^2 \{ \lambda_{ef}, \lambda_{gh} \} \Bigr] \Bigr\},
\\
&{\rm tr}_{\rm (s)} \Bigr\{ 
\lambda_{ab} \psi_c (D_d \lambda_{ef}) Q \lambda_{gh} \Bigr\} = 
{\rm tr} \Bigr\{ 
\lambda_{ab} \Bigr[ \psi_c (D_d \lambda_{ef}) Q \lambda_{gh} +
(Q \lambda_{gh} ) \psi_c D_d \lambda_{ef} -
(D_d \lambda_{ef} ) (Q \lambda_{gh}) \psi_c
\\
&\phantom{{\rm tr}_{\rm (s)} \Bigr\{ \lambda_{ab} \psi_c (D_d \lambda_{ef})
Q \lambda_{gh} \Bigr\} = {\rm tr} \Bigr\{ }
- (D_d \lambda_{ef}) \psi_c Q \lambda_{gh} +
\psi_c (Q \lambda_{gh}) D_d \lambda_{ef} -
(Q \lambda_{gh} ) (D_d \lambda_{ef}) \psi_c \Bigr] \Bigr\}.
\end{align*}

In order to pick out from (C.9) all the terms belonging to the minimal
sector, we have to evaluate the higher rank objects $\lambda_{ab}$,
$\lambda_{abc}, \ldots$ in (C.8) explicitly. By making use of (C.3)
after a simple calculation one obtains
\begin{align}
&\lambda_{ab} = \chi_{ab},
\nonumber
\\
&\lambda_{abc} = \Phi_{abc}^{~~~~\!m} D_m \bar{\phi},
\nonumber
\\
&\lambda_{abcd} = - \Phi_{abcd} [ \eta, \bar{\phi} ] -
\Phi_{abc}^{~~~~\!m} [ \chi_{dm}, \bar{\phi} ],
\nonumber
\\
&\lambda_{abcde} = - 5 \Phi_{abcd} [ D_e \bar{\phi}, \bar{\phi} ],
\nonumber
\\
&\lambda_{abcdef} = - 5 \Phi_{abcd}
[ [ \chi_{ef}, \bar{\phi} ], \bar{\phi} ],
\nonumber
\\
&\lambda_{abcdefg} = - 5 \Phi_{abcd} \Phi_{efg}^{~~~~\!m}
[ [ D_m \bar{\phi}, \bar{\phi} ], \bar{\phi} ],
\nonumber
\\
&\lambda_{abcdefgh} = 5 \Phi_{abcd} \Phi_{efgh}
[ [ [ \eta, \bar{\phi} ], \bar{\phi} ], \bar{\phi} ] +
5 \Phi_{abcd} \Phi_{efg}^{~~~~\!m}
[ [ [ \chi_{hm}, \bar{\phi} ], \bar{\phi} ], \bar{\phi} ],
\tag{C.10}
\\
\intertext{with}
&Q \lambda_{ab} = B_{ab},
\nonumber
\\
&Q \lambda_{abc} = \Phi_{abc}^{~~~~\!m} ( [ \psi_m, \bar{\phi} ] +
D_m \eta ),
\nonumber
\\
&Q \lambda_{abcd} = - \Phi_{abcd} ( [ [ \bar{\phi}, \phi ], \bar{\phi} ] -
\{ \eta, \eta \} ) - \Phi_{abc}^{~~~~\!m} ( [ B_{dm}, \bar{\phi} ] -
\{ \chi_{dm}, \eta \} ),
\tag{C.11}
\end{align}
where we have carried out similar replacements as in (C.7) and omitted
all the terms which do not contribute to (C.9) (remind that only
the fully antisymmetric part of $\lambda_{ab}$, $\lambda_{abc}, \ldots$
enters).

Then, by inserting into (C.9) for $\lambda_{ab}$, $\lambda_{abc}, \ldots$ the
expressions (C.10) and (C.11) and taking into account (\ref{4.7}), together
with the following basic identity \cite{15},
\begin{equation*}
\Phi^{abcm} \Phi_{defm} =
\delta^{[a}_d \delta^b_e \delta^{c]}_f +
\hbox{$\frac{1}{4}$} \Phi^{[ab}_{~~~[de} \delta^{c]}_{f]},
\end{equation*}
after a tedious calculation we could express $\Psi_{\rm ext}$ in terms of
$A_a$, $\psi_a$, $\chi_{ab}$, $\eta$, $\phi$, $\bar{\phi}$ and $B_{ab}$.
In order to select from $\Psi_{\rm ext}$ the terms belonging to the minimal
sector, we rescale $\chi_{ab}$, $\eta$, $\bar{\phi}$ and $B_{ab}$ as well as
$\Psi_{\rm ext}$ with the gauge parameter $\xi$ and $1/\xi$, respectively.
Then, by putting $\xi$ equal to zero, i.e., by choosing the Landau type gauge,
$\Psi_{\rm ext}$ considerably simplifies into
\begin{equation}
\tag{C.12}
\Psi_{\rm ext}^\prime = \hbox{$\frac{1}{8!}$} \epsilon^{abcdefgh}
\int_E d^8x\, {\rm tr}_{\rm (s)} \Bigr\{
2520 \lambda_{ab} F_{cd} F_{ef} F_{gh} -
1680 \lambda_{abc} \psi_d F_{ef} F_{gh} \Bigr\}.
\end{equation}
Moreover, it is easily seen that the further evaluation of the right--hand
side of (C.12) reveals precisely the expression (\ref{3.5}) we are looking
for.

\end{document}